\documentclass[12pt]{article}
\usepackage{amsfonts,floatflt}
\usepackage{epsfig}

 \def\beq#1{\begin{equation}\label{#1}}
 \def\eeq{\end{equation}}

 \newcommand{\bear}[1]{\begin{eqnarray}\label{#1}}
 \newcommand{\ear}{\end{eqnarray}}

  \renewcommand{\theequation}{\arabic{section}.\arabic{equation}}
 \catcode`\@=11 \@addtoreset{equation}{section}\catcode`\@=12

 \newcommand{\R}{ {\mathbb R} }

 \newcommand{\sign}{\mathop{\rm sign}\nolimits}
 \newcommand{\eps}{\varepsilon}
 
 \newcommand{\nn}{\nonumber}

 \newcommand{\fnm}{\footnotemark}
 \newcommand{\fnt}{\footnotetext}

 \def\noi{\noindent}
 \def\dys{\displaystyle}
 \def\mst{\mathstrut}

 \vspace{15pt}

\begin{document}

  \begin{center}
  \large\bf

 On the ``scattering law'' for Kasner parameters
 in the model with one-component  anisotropic fluid

 \vspace{15pt}

 \normalsize\bf V.D. Ivashchuk\fnm[1]\fnt[1]{ivashchuk@mail.ru}
             and  V.N. Melnikov\fnm[2]\fnt[2]{melnikov@vniims.ru}

 \vspace{5pt}

 \it Center for Gravitation and Fundamental Metrology,
 VNIIMS, 46 Ozyornaya ul., Moscow 119361, Russia  \\

 Institute of Gravitation and Cosmology,
 Peoples' Friendship University of Russia,
 6 Miklukho-Maklaya ul.,  Moscow 117198, Russia \\

 \end{center}

 \vspace{15pt}

 \small\noi

 \begin{abstract}
    A multidimensional cosmological type model
    with 1-component  anisotropic fluid
    is considered.  An exact solution
     is obtained. This solution is defined on
    a product manifold  containing  $n$ Ricci-flat factor spaces.
      We singled out a special solution  governed
    by the function $cosh$. It is shown that this special solution
    has  Kasner-like asymptotics in the  limits $\tau \to  + 0$ and
    $\tau \to  + \infty$,  where $\tau$ is a synchronous time variable.
    A relation between two sets of Kasner parameters
    $\alpha_{\infty}$ and  $\alpha_{0}$ is found.
    This formula (of ``scattering law'') is coinciding with that
    obtained earlier for the $S$-brane solution
    (when scalar fields are absent).
 \end{abstract}

 \vspace{20cm}

 \pagebreak

 \normalsize

\section{Introduction}

  In this paper  we continue our investigations
  (started in  \cite{IM-sl-09}) of
  multidimensional solutions defined  on product of
  several Ricci-flat factor spaces which have two Kasner-like
  asymptotical regions.

 Here we recall that Kasner-like solutions with a chain of
 $n$ Ricci-flat factor-spaces
 $(M_i,g^{(i)})$  have the following form \cite{I-92}
 \bear{1.m}
  g = w d\tau \otimes d\tau + \sum_{i=1}^{n} A_i^2
                           \tau^{2 \alpha^i} g^{(i)},
  \ear
  where $w = \pm 1$, $\tau > 0$,
  \bear{1.k1}
   \sum_{i=1}^{n} d_i \alpha^i = 1,
   \\  \label{1.k2}
  \sum_{i=1}^{n} d_i (\alpha^i)^2  = 1, \
  \ear
  and for any $i = 1, \ldots, n$ ($n \geq 2$):
  $A_i > 0$ is constant, $g^{(i)}$ is a Ricci-flat
  metric defined on the manifold $M_i$
    (for $w = -1$ see \cite{I-92}).

  These solutions with non-Milne-type
  sets of Kasner parameters are singular since the
  Riemann tensor squared is divergent as $\tau \to + 0$
  \cite{IM-95a}. For Milne-type sets of parameters, i.e. when $d_i = 1$
  and $\alpha^{i} = 1$ for some $i$ ($\alpha^j = 0$ for
  all $j \neq i$)   the metric is regular as $\tau \to + 0$, when
  either i) $g^{(i)} = - w dy^i \otimes dy^i$, $M_i = \R$
  ($ - \infty < y^i < + \infty$),
  or ii) $g^{(i)} = w dy^i \otimes dy^i$, $M_i$
  is circle of length $L_i$ ($0 < y^i < L_i$)
  and $A_i L_i = 2 \pi$ (i.e. when the cone singularity is absent).

   In this paper  we consider an exact cosmological type solution with
   1-component ``perfect'' fluid (Section 2).
   (For earlier publications on multidimensional cosmological models
   with perfect fluid see \cite{Sah}-\cite{IM-95}
   and references therein.)  This solution is defined on a
   product manifold  containing  $n$ Ricci-flat factor spaces.
   It is derived in the Appendix.  For $w = -1$ it was found in
   \cite{IM-89,IM-94,GIM-95} and generalized in
   \cite{IM-95} for the case when a scalar field was added.
    A special case of this solution with a $\Lambda$-term
    component  was obtained in \cite{IM-tmf-94} (see also
    \cite{BIMZ-94} for scalar field generalization).

    We  write the solution in   a so-called ``minisuperspace-covariant''
   form that significantly simplifies the forthcoming analysis.
   In Section 3 we single out a special solution   governed  by the $cosh$ function.
   We show that this solution has a Kasner-like asymptotics in
    both limits $\tau \to  + 0$ and $\tau \to  + \infty$,
   where $\tau$ is the synchronous time variable.
   We also find a relation between two sets of Kasner parameters
   $\alpha_{\infty} = (\alpha_{\infty}^{i}) \in \R^n$
   and  $\alpha_0 = (\alpha_{0}^{i}) \in \R^n$:
     \beq{gcl}
    \alpha_{\infty}^{i} =
               \frac{\alpha_0^i - 2 U(\alpha_0) U^i (U,U)^{-1}}
               {1 - 2 U(\alpha_0) (U,U^{\Lambda})(U,U)^{-1}},
  \eeq
  $i = 1, \dots, n$.  Here  $U = (U_i)$ is a  co-vector
  corresponding to the fluid component, $\bar{U} =(U^i)$ is dual vector and
  $U^{\Lambda}$  is a co-vector, corresponding to the $\Lambda$-term.
  All  these vectors and the scalar product $(.,.)$  are defined below
  (see Section 2). Here $U(\alpha_0) = U_i \alpha_0^i > 0$ and
   $U(\alpha_{\infty}) = U_i \alpha_{\infty}^i < 0$.

    A  relation analogous to (\ref{gcl}) (``scattering law'' formula) was obtained
    earlier  for S-brane solution with one brane in
    \cite{IM-sl-09}.  We note that in \cite{IM-sl-09}
    the geometrical sense   of the scattering law was clarified for $n > 2$.
    Namely,  the scattering law transformation
    for a brane $U$-vector  (obeying $(U,U^{\Lambda}) < 0$)
    was expressed  in terms of a function  mapping a ``shadow'' part of the
    Kasner sphere $S^{n-2}$ onto  ``illuminated'' one.
    The shadow  and illuminated parts of the Kasner sphere were defined
    w.r.t.  a point-like source of light located
    outside the Kasner sphere $S^{n-2}$. (For details of this
    geometrical construction see \cite{IM-sl-09}).

      The relation (\ref{gcl}) appears also  when the billiard approach
   to multicomponent anisotropic fluid is considered
   \cite{IKM-94a,IKM-94b,IM-bil-95,IMRC}.  It may be shown
  (as it was done in \cite{DH,Ierice}   for $S$-brane solutions)  that
  after the collision with a billiard  wall
  (corresponding to the fluid component) the set
  of Kasner parameters, is defined by the Kasner set before
  the collision   through the formula analogous to (\ref{gcl}), see
  \cite{IM-bil-rev09}.  For the  billiard approach in models with scalar field and fields of
  forms see \cite{IMb1,DH,DHN,IM-bil-rev09} and refs. therein.

\section{Model with anisotropic fluid and its exact solution}

\subsection{The set-up}

 Now, we consider a  cosmological type solution to Einstein equations
with an anisotropic (perfect) fluid  matter source

 \beq{1.1}
  R^M_N - \frac{1}{2}\delta^M_N R = k^2 T^M_N
 \eeq
 defined on $D$-dimensional manifold
  \beq{1.2}
    M = {\R}_{.}  \times M_1 \times M_2 \times \ldots \times M_n,
    \eeq
 with  block-diagonal metric
 \beq{1.3a}
  g = w e^{2\gamma (u)}
  du \otimes du + \sum_{i=1}^{n} e^{2 \beta^i(u)} g^{(i)}.
 \eeq

 Here $\R_{.} = (u_{-},u_{+})$ is an interval,  $w = \pm1$ and $n \geq 2$.
 Manifold $M_i$ with the metric $g^{(i)}$ is a Ricci-flat space
 of dimension $d_{i}$: $R_{m_{i}n_{i}}[g^{(i)}]=0$,
  $i=1,2,\ldots,n$,  and $\kappa^2$ is a multidimensional
  gravitational constant.

Energy-momentum tensor of anisotropic fluid is adopted in the
following form:
  \beq{1.5}
 (T^{M}_{N})= {\rm diag}(-\hat{\rho},{\hat p}_{1} \delta^{m_{1}}_{k_{1}},\ldots
 , {\hat p}_n \delta^{m_{n}}_{k_{n}}),
 \eeq
where $\hat{\rho}$ and $\hat p_{i}$ are  ``density'' and
``pressures'', respectively, depending upon  radial variable $u$.

In the cosmological case when $w = -1$ and all metrics $g^{(i)}$
have  Euclidean signatures,   $\hat{\rho} = \rho$ is a density and
${\hat p}_i = p_i$ is a pressure in $i$-th space. For static
configurations with $w = 1$,  $g^{(1)} = - dt \otimes dt$ and all
metrics $g^{(i)}$, $i > 1$,  having Euclidean signatures,
 the physical density and pressures are related to the effective
(``hat'') ones by formulas:
  $\rho = - {\hat p}_1$, $p_u = - \hat{\rho}$,
  $p_i = \hat{p}_i, \quad (i \neq 1)$, where $p_u$ is the
  pressure in $u$-th direction.

 We also impose the following equation of state
 \beq{1.7}
 {\hat p}_i= \left(1-\frac{2U_i}{d_i}\right){\hat{\rho}},
 \eeq
 where $U_i$ are constants, $i= 1,2,\ldots,n$.

 In what follows  we use a scalar product
  \beq{2.4}
  (U,U')= G^{ij} U_i U'_j = \sum_{i = 1}^{n} \frac{U_i U'_i}{d_i} +
  \frac{1}{2 -D} (\sum_{i = 1}^{n} U_i) (\sum_{j = 1}^{n} U'_j),
  \eeq
 for $U = (U_i), U' = (U'_i) \in \R^n$,
 where
 \beq{2.2b}
    G^{ij}=\frac{\delta^{ij}}{d_i} + \frac1{2-D}
 \eeq
 are components of dual minisuperspace metric.
 Recall that $(G^{ij}) = (G_{ij})^{-1}$, where
 \beq{2.2a}
   G_{ij}= d_i \delta_{ij} - d_i d_j,
 \eeq
 are components of minisuperspace metric \cite{IMZ}.

   We also define a co-vector
  \beq{2.4l}
   U^{\Lambda} = (d_i),
  \eeq
   corresponding to the $\Lambda$-term
  and the vector $\bar{U} =(U^i)$
  \beq {1.16}
  U^{i} = G^{ij} U_j  = \frac{U_i}{d_i} + \frac{1}{2 -D} \sum_{j = 1}^{n} U_j,
  \eeq
  which is dual to $U$.

 \subsection{Exact solution }

 Here we consider an exact cosmological solution to
 Hilbert-Einstein equations  (\ref{1.1})
 defined  on the manifold (\ref{1.2}). We impose
 the following restriction on the $U$-vector in
 (\ref{1.7})

  \beq{1.17}
  K = (U,U) = \sum_{i = 1}^{n} \frac{U_i^2}{d_i} +
  \frac{1}{2 -D} (\sum_{i = 1}^{n} U_i)^2 \neq 0.
 \eeq

 (The case $K$ = 0 will be considered in a separate publication.)

  The solution has the following form (see Appendix C)
 \bear{1.3}
  g=   |f(u)|^{- 2 h (U,U^{\Lambda})}
   \exp(2c^0 u + 2 \bar c^0) w du \otimes du  + \\ \nn
   \sum_{i=1}^{n} |f(u)|^{- 2 h  U^i}
       \exp(2c^i u+ 2 \bar c^i) g^{(i)} ,
  \\ \label{1.4}
   k^2 \hat{\rho} = - w A |f(u)|^{2 h (U,U^{\Lambda}) -2}
   \exp(- 2c^0 u - 2 \bar c^0),
  \ear
  where  $w = \pm 1$,  $h = K^{-1}$,
  $g^{(i)}$  is a Ricci-flat  metric on $M_{i}$,
  and
 \beq{1.16a}
 (U,U^{\Lambda}) =  \frac{\sum_{i =1}^n U_i}{2 -D},
 \eeq
 $i=  1,\ldots,n$.

 The moduli function $f$ reads
 \bear{1.4.5}
  f(u)=
  R \sinh(\sqrt{C}(u-u_0)), \;
  C > 0, \; K A <0; \\ \label{1.4.7}
  R \sin(\sqrt{|C|}(u-u_0)), \;
  C<0, \; K A < 0; \\ \label{1.4.8}
  R \cosh(\sqrt{C}(u-u_0)), \;
  C > 0, \; K A > 0; \\ \label{1.4.9}
  |2A K|^{1/2}(u-u_0), \; C=0, \; K A <0,
  \ear
 where $R = |2A K/ C|^{1/2}$, and $C$, $u_0$  are constants.
 (In  (\ref{1.3}) and (\ref{1.4})  $f(u) \neq 0$  is assumed
  for all $u \in (u_{-},u_{+})$.)

 Vectors $c= (c^i)$ and
 $\bar c=(\bar c^i)$ obey the following constraints:
 \bear{1.27}
   U(c) =  U_i c^i =0,   \qquad   U(c) = U_i \bar c^i = 0
   \\  \label{1.30a}
   C  K^{-1} + G_{ij} c^i c^j  =0,
  \ear
   where $G_{ij} c^i c^j =  \sum_{i=1}^n d_i(c^i)^2
                     - (\sum_{i=1}^n d_ic^i)^2$.

    In  (\ref{1.3}) and (\ref{1.4}) we also denote
    \beq{2.c}
     c^0 =  U^{\Lambda}(c) =\sum_{i=1}^n d_i c^i, \quad
      \bar c^0 = U^{\Lambda}(\bar c) = \sum_{i=1}^n d_i \bar c^i.
   \eeq

  The special solution with $C = c_i = 0$ (for all $i$) and $w = -1$ was considered
  in detail in \cite{AIKM,IKMN}. For $U = U^{\Lambda}$ and $A >0$ it contains
  a special solution with $d_i = 1$, $g^i = dy^i \otimes dy^i$ ($i = 1, ..., n$),
  describing either (a part of) de-Sitter space (for $w = -1$) or
  (a part of) anti-de-Sitter space (for $w = 1$).

   {\bf Minisuperspace-covariant form of solution.}

   This solution is derived in Appendix C in terms of
   ``minisuperspace-covariant'' notations for
   functions $\gamma(u)$, $\beta^i(u)$ appearing
   in  metric (\ref{1.3a}).

  Solution for  $\beta = (\beta^{i}(u))$ reads
   as follows:
  \beq{2.0}
   \beta^i(u)= - \frac{U^{i}}{(U,U)}\ln |f(u)| + c^i u +  \bar{c}^i,
  \eeq
  where  $f(u)$ was defined in
  (\ref{1.4.5})-(\ref{1.4.9}) and
 \beq{2.gd}
  \gamma = \gamma_0 \equiv \sum_{i=1}^n d_i \beta^i =  U^{\Lambda}_i \beta^i
 \eeq
 and   $u$ is the harmonic variable.

 \section{Scattering law for Kasner parameters}

  Now we restrict our consideration by a special solution
  with  $C >0$, $K  = (U,U) > 0$  and $A > 0$.
  In this case the solution  is governed  by
  moduli function  $f(u)= R \cosh(\sqrt{C}(u- u_0))$,
  $u \in (- \infty, + \infty)$,  and
  has  two Kasner-like asymptotics in the  limits $\tau \to  + 0$ and
   $\tau \to  + \infty$,  where $\tau$ is a synchronous time variable
  (see below).

  Another case, when there are two Kasner-like asymptotical
  regions,  takes place when $C >0$, $K  = (U,U) < 0$ and $A < 0$ (this will
  be a subject of  a separate paper).

  \subsection{Kasner-like behaviour}

 Let us consider our solution in a synchronous time:
 \beq{3.tau}
  \tau = \eps  \int ^{u}_{u_0} d \bar{u} e^{\gamma_0( \bar{u})},
 \eeq
 where $\eps = \pm 1$,  and
  \beq{3.l}
   e^{\gamma_0(u)} = |f(u)|^{- h (U^{\Lambda},U)} \exp(c^0 u  + \bar c^0)
  \eeq
 is a lapse function.

 Due to
 \beq{3.fas}
   f \sim \frac{R}{2} \exp( \pm \sqrt{C}(u - u_0)),
 \eeq
  for $u \to \pm \infty$,
  we get asymptotical relations for the lapse function
  \beq{3.flas}
   e^{\gamma_0} \sim {\rm const} \exp( b_{\pm} \sqrt{C} u ),
  \eeq
  as $u \to \pm \infty$,  with
  \beq{3.1}
    b_{\pm} = \mp h (U^{\Lambda},U) +  \frac{c^0}{\sqrt{C}}.
  \eeq

  Using relations (\ref{2.c})  and $h = (U,U)^{-1}$,
  we could rewrite parameters  $b_{\pm}$ in
  a minisuperspace-covariant form:
   \beq{3.b}
     b_{\pm} = \mp \frac{(U^{\Lambda},U)}{(U,U)} +  (s,U^{\Lambda}),
   \eeq
  where
  \beq{3.s}
    s  = (s_i) = (G_{ij} c^j/\sqrt{C})
  \eeq
  is a co-vector, obeying relations
   \bear{3.2}
       (s,U) = 0,
   \\ \label{3.3}
     \frac{1}{(U,U)}  + (s,s) = 0,
  \ear
  following just from  (\ref{1.27}) and (\ref{1.30a}).  In
  derivation of (\ref{3.b}) we used the relation
   \beq{3.c}
    c^0  = (s,U^{\Lambda}) \sqrt{C},
   \eeq
   following from (\ref{2.c}) and (\ref{3.s}).

   In what follows we will use the inequality
   \beq{3.4}
    |(s,U^{\Lambda})| >  \frac{|(U^{\Lambda},U)|}{(U,U)},
   \eeq
    proved in Appendix C. The proof used relations
   (\ref{3.2}), (\ref{3.3}) and $(U,U) > 0$.

   The parameter $c^0$ is a non-zero one (otherwise the relation (\ref{1.30a})
   would be incompatible with the conditions $C >0$, $K > 0$).

   It follows from (\ref{3.4}) that $b_{\pm}$
   are also non-zero and
    \beq{3.5}
    \sign(b_{\pm}) = \sign((s,U^{\Lambda})) = \sign(c^0).
   \eeq

   It may be verified that due to (\ref{3.4}) the lapse function $e^{\gamma_0(u)}$
   is monotonically increasing from $+0$ to $+ \infty$ for $c^0 >0$
   and monotonically decreasing from $+ \infty$ to $+0$ for $c^0 <
   0$.

     We define a synchronous-like variable to be
     \beq{3.t1}
      \tau =   \int^{u}_{- \infty} d \bar{u} e^{\gamma_0( \bar{u})}
     \eeq
   for $c^0 >0$ and
     \beq{3.t2}
      \tau =   \int^{+ \infty}_{u} d \bar{u} e^{\gamma_0( \bar{u})}
     \eeq
   for $c^0 < 0$. Then,  $\tau = \tau(u)$
   is monotonically increasing from $+0$ to $+ \infty$ for $c^0 >0$
   and monotonically decreasing from $+ \infty$ to $+0$ for $c^0 <
   0$.

     We have the following asymptotical relations for $\tau = \tau(u)$
    \beq{3.6}
    \tau \sim {\rm const} \  b_{\pm}^{-1} \exp( b_{\pm} \sqrt{C} u ),
    \eeq
    as $u \to \pm \infty$.

    For     $\beta = (\beta^i)$ from (\ref{2.0})  we get (see (\ref{3.fas}))
   \beq{3.7}
    \beta^i(u) \sim  \mp \frac{U^{i} \sqrt{C} u}{(U,U)} + c^i u +  \hat{c}^i
   \eeq
   as $u \to \pm \infty$, where $\hat{c}^i$ are constants.
   Hence, due to (\ref{3.6}), we are led to Kasner-like
   asymptotics
    \beq{3.8}
     \beta^i \sim  \alpha^i_{\pm} \ln \tau +  \beta^i_{\pm}
    \eeq
     for $u \to \pm \infty$, where $\beta^i_{\pm}$ are constants and
    \beq{3.9}
       \alpha^i_{\pm} =  [\mp \frac{U^{i}}{(U,U)} +  s^{i}]/b_{\pm}
    \eeq
    are  Kasner-like parameters  corresponding to $u \to \pm \infty$.

    Asymptotical relations (\ref{3.8}) could be also rewritten
    in the form of proper time asymptotics, i.e.
    \bear{3.10a}
     \beta^i \sim  \alpha^i_{0} \ln \tau +  \beta^i_0, \ {\rm as} \ \tau \to
     +0,    \\ \label{3.10b}
     \beta^i \sim  \alpha^i_{\infty} \ln \tau +  \beta^i_{\infty}, \ {\rm as} \ \tau \to
     + \infty.
    \ear
     Here
     \beq{3.11a}
      \alpha^i_{0} = \alpha^i_{-}, \quad \alpha^i_{\infty} = \alpha^i_{+}
     \eeq
   for $c^0 >0$ and
      \beq{3.11b}
      \alpha^i_{0} = \alpha^i_{+}, \quad \alpha^i_{\infty} = \alpha^i_{-}
     \eeq
     for $c^0 < 0$ and $\beta^i_{0}$, $\beta^i_{\infty}$ are constants.

  It follows from definitions of Kasner parameters (\ref{3.9}) that
   \bear{3.12}
     G_{ij}\alpha^i_{\pm} \alpha^j_{\pm}   = 0, \\
    \label{3.13}
    U(\alpha_{\pm}) = U_i \alpha^i_{\pm}  = \mp \frac{1}{b_{\pm}}, \\
    \label{3.14}
    U^{\Lambda}(\alpha_{\pm})  =  1,
    \ear
    see (\ref{3.b}), (\ref{3.2}) and (\ref{3.3}).

   In components relations (\ref{3.12}) and  (\ref{3.14})
   read as
   \beq{3.15k1}
    \sum_{i=1}^{n} d_i \alpha^i_{\pm} =
    \sum_{i=1}^{n} d_i (\alpha^i_{\pm})^2 = 1. \
    \eeq

    Thus, we are led to Kasner-like relations (\ref{1.k1}) and
    (\ref{1.k2})
    for $\alpha_{\pm} = (\alpha^i_{\pm})$.
    Hence, $\alpha_{0} = (\alpha^i_{0})$ and $\alpha_{\infty} = (\alpha^i_{\infty})$
    also obey relations (\ref{1.k1}) and (\ref{1.k2}).

    So, we
    obtained a Kasner-like asymptotical behaviour of our special solution
    (with $C >0$, $K > 0$ and $A > 0$)  for i) $\tau \to +0$ and for
    ii) $\tau \to   + \infty$, as well.  The Kasner-like  behaviour in the case i)
    is in agreement with the general result of the billiard approach from \cite{IMb1}.
    The  the case ii) was considered in \cite{IM-bil-rev09}.

    Using  (\ref{3.5}) and (\ref{3.13}) we get
       \bear{3.16i}
       U(\alpha_{0}) = U_i  \alpha^i_{0}  > 0,
       \\ \label{3.16ii}
       U(\alpha_{ \infty})    < 0.
       \ear

       \subsection{Scattering law}

  Now, we derive a relation between Kasner sets  $\alpha_{0}$
  and  $\alpha_{\infty}$.

   We start with formulae:
    \beq{3.17}
     b_{+} \alpha_{+} - b_{-} \alpha_{-} =  -  \frac{2 \bar{U}}{(U,U)}
    \eeq
   and
   \beq{3.18}
     b_{+}  - b_{-} =  -  \frac{2(U^{\Lambda},U)}{(U,U)},
    \eeq
   following from (\ref{3.9}) and  (\ref{3.b}), respectively.
   (Recall that $\bar{U} = (U^i)$. )
    Using these relations  and (\ref{3.13})  we get
    \beq{3.19}
      \alpha_{\pm}^i =
               \frac{\alpha_{\mp}^i - 2 U^i U(\alpha_{\mp}) (U,U)^{-1}}
               {1 - 2 U(\alpha_{\mp}) (U,U^{\Lambda})(U,U)^{-1}} .
     \eeq
     This formula gives a scattering law formula for Kasner parameters
     in our case (see definitions (\ref{1.16}), (\ref{3.11a}) and (\ref{3.11b}))
     or
    \beq{3.20}
      \alpha_{ \infty} =
               \frac{\alpha_{ 0} - 2 \bar{U} U(\alpha_{ 0}) (U,U)^{-1}}
               {1 - 2 U(\alpha_{ 0}) (U,U^{\Lambda})(U,U)^{-1}} = S(\alpha_0).
     \eeq
    coinciding  with the scattering law formula (\ref{gcl})
     derived in \cite{IM-sl-09} for another $S$-brane solution when scalar fields are
     absent and $U$ is coinciding with the brane $U$-vector.

    Due to (\ref{3.19})  the inverse function $S^{-1}$
    is given by just the same relation
     \beq{3.20a}
      \alpha_{0} =
               \frac{\alpha_{\infty} - 2 \bar{U} U(\alpha_{\infty}) (U,U)^{-1}}
               {1 - 2 U(\alpha_{\infty }) (U,U^{\Lambda})(U,U)^{-1}}
               = S^{-1}(\alpha_{\infty}).
     \eeq

 \subsection{Geometric meaning of the scattering law}

 Here we analyze the geometric meaning of the scattering
 for $n > 2$ as it was done in  \cite{IM-sl-09} for
 the $S$-brane solution.

  The Kasner-like relations (\ref{1.k1}) and (\ref{1.k2}) describe
 an ellipsoid isomorphic to a unit $(n-2)$-dimensional sphere
 $S^{n-2}$ belonging to  $\R^{n-1}$.
 The sets of Kasner parameters $\alpha$
 may be parametrized  by vectors $\vec{n}  \in S^{n-2}$, i.e.
  $\alpha= \alpha(\vec{n})$.

   For $(U,U^{\Lambda}) \neq 0$
   (or, equivalently, when  $\sum_{i =1}^n U_i \neq 0$, see (\ref{1.16a}))
   the scattering law formula (\ref{gcl}) in terms
   of $\vec{n}$-vectors reads as in \cite{IM-sl-09}
   \beq{4.gcl}
   \vec{n}_{\infty} = \frac{(\vec{v}^2 - 1) \vec{n}_0
   + 2(1 - \vec{v}\vec{n}_0)\vec{v}}{(\vec{v} - \vec{n}_0)^2}
   \eeq
   where $\vec{v}$ is a vector belonging to $\R^{n-1}$ with
   $|\vec{v}| > 1$.

    Here
    \beq{4.2a}
      \vec{v} \vec{n}_{0} < 1 \qquad \vec{v} \vec{n}_{\infty} > 1,
     \eeq
    for $(U,U^{\Lambda}) < 0$ (or, equivalently, when  $\sum_{i =1}^n U_i > 0$)
    and
    \beq{4.2b}
      \vec{v} \vec{n}_{0} > 1 \qquad \vec{v} \vec{n}_{\infty} < 1,
     \eeq
    for $(U,U^{\Lambda}) > 0$ (or, equivalently, when  $\sum_{i =1}^n U_i <
    0$).

    The vector    $\vec{v} = (v_i) \in \R^{n-1}$
    is defined by the formula
    \beq{4.19}
      v_i  = - \hat{U}_i/\hat{U}_0,
    \eeq
    $i = 1, \ldots, n-1$, where
    \beq{4.11}
    \hat{U}_a  = e_{a}^{i} U_i,
    \eeq
    and the invertible matrix  $(e^{a}_{i})$  satisfies the relations
   \beq{4.5}
   \eta^{ab} =
   e^{a}_{i} G^{ij} e^{b}_{j},
   \eeq
   $a,b = 0, \ldots, n-1$, with
   \beq{4.10}
    e^{0}_i = q^{-1} U^{\Lambda}_i,
   \eeq
  and
   \beq{4.10q}
   q =  [- (U^\Lambda,U^\Lambda)]^{1/2} = [(D-1)/(D-2)]^{1/2}.
   \eeq
   (Here   $(\eta_{ab})=(\eta^{ab}) = diag(-1,+1, \ldots,+1)$.)

   This implies
  \beq{4.12}
   \hat{U}_0  =  - q^{-1} (U,U^{\Lambda})
  \eeq
   and hence $\hat{U}_0 \neq 0$ when $(U,U^{\Lambda}) \neq 0$.

  Relations (\ref{4.gcl}), (\ref{4.2a}) and (\ref{4.2b})
  could be readily proved
  from (\ref{3.19}), (\ref{3.16i}) and (\ref{3.16ii})
  if  the following ``frame'' Kasner-like parameters
  \beq{4.15}
  \hat{\alpha}^a  = e^a_{i} \alpha^i,
  \eeq
  with
     \beq{4.18}
   \hat{\alpha}^0 = q^{-1}, \qquad \hat{\alpha}^i = q^{-1} n^i,
   \eeq
   $i = 1, \ldots, n-1$, are used (see \cite{IM-sl-09}).
   An important relation here is the following one
   \beq{4.21}
      U(\alpha) = U_A \alpha^A = \hat{U}_a \hat{\alpha}^a =
      q^{-1} \hat{U}_0 (1 - \vec{v}\vec{n}).
    \eeq

    Thus, for  $(U,U^{\Lambda}) \neq 0$
    we get just a modified inversion with respect to a point $v$
    located outside the Kasner sphere $S^{n-2}$ (see Fig. 1).
    For  $(U,U^{\Lambda}) < 0$
    the function  (\ref{4.gcl}) maps  a shadow part of the
    Kasner sphere $S^{n-2}$ onto  illuminated one, while
    for  $(U,U^{\Lambda}) > 0$
    this function  maps  an illuminated  part of the
    Kasner sphere $S^{n-2}$ onto shadow  one.
     Here the shadow
    and illuminated parts of the Kasner sphere are defined
    w.r.t.  a point-like source of light located at $v$.

    \begin{floatingfigure}[l]{6.25 cm}
    \includegraphics[height=6 cm,width=6cm, keepaspectratio]{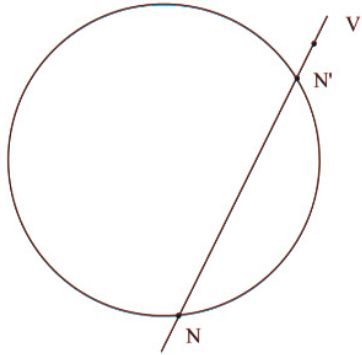}
    \caption{\textit{The graphical representation of the
    modified inversion $S$ w.r.t. a point $V$
    for $n = 3$, and $(U,U^{\Lambda}) < 0$:
    $N'= S(N)$.}}
    \end{floatingfigure}

     For $(U,U^{\Lambda}) = 0$
    (or, equivalently, when  $\sum_{i =1}^n U_i = 0$)
    the main formula (\ref{gcl}) in terms
    of $\vec{n}$-vectors reads
    \beq{4.gcl-0}
    \vec{n}_{\infty} = \vec{n}_0  - 2( \vec{b}\vec{n}_0)\vec{b},
    \eeq
    where $\vec{b} = (b_i)$ is a unit vector belonging to $\R^{n-1}$ (
    $|\vec{b}| = 1$) with components
    \beq{4.19b}
      b_i  =
      \hat{U}_i/(\sum_{j =1}^{n-1} \hat{U}_j^2)^{1/2},
    \eeq
    $i = 1, \ldots, n-1$. The inequalities on Kasner-like
    parameters (\ref{3.16i}) and (\ref{3.16ii}) in this case
    reads as follows
     \beq{4.2c}
      \vec{b} \vec{n}_{0} >  0, \qquad \vec{b} \vec{n}_{\infty} < 0.
     \eeq

     Thus, for   $(U,U^{\Lambda}) =0$  the function  (\ref{4.gcl})
     is just a reflection with respect to  a hyperplane
     $\{\vec{y}:  \vec{b} \vec{y} =0 \}$,
     which    contains a center of the Kasner sphere.

     Relations (\ref{4.19b}) and (\ref{4.2c}) may be obtained
     from (\ref{4.gcl}), (\ref{4.2a}) and  (\ref{4.2b}) by means
     of the limiting procedure: $\hat{U}_0 \to \pm 0$ ($|\vec{v}| \to + \infty $.).

     It should be noted that all formulas presented above are also valid for
     $n = 2$. In this case the zero-dimensional Kasner sphere $S^0 = \{-1, 1\}$ should
     be considered.

   \section{Example: $n = 2$}

   Here we consider the simplest case of the solution with $C>0$, $K > 0$, when $n
   =2$. We put $U_1 \neq 0$ and $U_2 = 0$,
   i.e. ${\hat p}_1= w_1 \hat{\rho}$ with $w_1 \neq 1$ and ${\hat p}_2=
   \hat{\rho}$.

   For Kasner set $\alpha = ( \alpha^1, \alpha^2)$ we get from
   (\ref{1.k1}) and (\ref{1.k2}) \cite{BIMZ-94,GIM-96}
   \beq{5.1}
    \alpha_{\pm} = (\alpha_{\pm}^1, \alpha_{\pm}^2)
   = \frac{1}{d_1 + d_2} \left(1 \pm \frac{r}{d_1}, 1 \mp \frac{r}{d_2}
   \right),
   \eeq
   where $r = \sqrt{d_1 d_2 (d_1 + d_2 -1)}$. (The number $r >0$ is
   integer one when $d_1 =1$ or $d_2 =1$ and also for
    $(d_1, d_2) = (3,6), (5,5), (2, 8), (13, 13)$ etc
   \cite{GIM-96}.)

   Let $d_2 > 1$. Then $\alpha_{+}^1 > 0$ and $\alpha_{-}^1 < 0$.
   Due to  $U_2 = 0$: $U(\alpha) = U_1 \alpha^1$ and
   hence $U(\alpha_{+}) > 0$ and $U(\alpha_{-}) < 0$ for $U_1 > 0$
   ($w_1 < 1$) and
   $U(\alpha_{+}) < 0$ and $U(\alpha_{-}) > 0$ for $U_1 < 0$ ($w_1 < 1$).

   It follows from (\ref{3.16i}) and (\ref{3.16ii})
   that
     \beq{5.2i}
       \alpha_{0} = \alpha_{+}, \qquad
       \alpha_{\infty} = \alpha_{-}.
       \eeq
   for $U_1 > 0$ and
   \beq{5.2iii}
       \alpha_{0} = \alpha_{-}, \qquad
       \alpha_{\infty} = \alpha_{+}.
       \eeq
  for $U_1 < 0$.

   Relation $U_i c^i = U_1 c^1 = 0$ implies $c^1 = 0$. Here $c^0 = d_2
   c^2$. Due to (\ref{3.11a}) and (\ref{3.11b}) we should put
   $c^2 < 0$ for $U_1 > 0$ and $c^2 > 0$ for $U_1 < 0$. In this
   case the sets $\alpha_{\pm}$ given by (\ref{3.9}) are coinciding
   with   those given by (\ref{5.1}). This may be also verified by
   straightforward calculations using the following relations

     \bear{5.3a}
       U^1 = \frac{(d_2 -1) U_1}{d_1 (D - 2)},
       \qquad  U^2 = \frac{U_1}{(2 - D)} = (U,U^{\Lambda}),
       \\ \label{5.3b}
       K = U^1 U_1, \qquad C = K d_2 (d_2 - 1) c_2^2,
        \ear
    where $D = d_1 + d_2 + 1 > 3$.

   {\bf Accelerated expansion of 3-dimensional factor-space.}
   After replacing $\tau \to \tau_0 - 0$, where $\tau_0$ is
   constant, we get for $w = -1$ two asymptotical
   Kasner type metrics

   \beq{5.a}
   g_{as} = - d\tau \otimes d\tau + \sum_{i=1}^{2} A_i^2
                           (\tau_0 - \tau)^{2 \alpha^i} g^{(i)},
  \eeq
   where either $\alpha^i = \alpha^i_0$
   ($A_i = A_{i,0} > 0$ )  as $\tau \to \tau_0 - 0$,
   or $\alpha^i = \alpha^i_{\infty}$
   ($A_i = A_{i, \infty } > 0$ ) as $\tau \to  - \infty$.

   Let $M_1$ be a flat 3-dimensional factor space ($d_1 = 3$),
   with the metric
   $g^{(1)} = dy^1 \otimes dy^1 + dy^2 \otimes dy^2 + dy^3 \otimes dy^3$.
   Then, due to relations (\ref{5.2i}), (\ref{5.2iii}) and
   $\alpha_{\infty} < 0$ for $d_2 > 1$, we get an asymptotical
   accelerated expansion of our 3-dimensional factor space $M_1$
   either as  $\tau \to \tau_0 - 0$ for $U_1 < 0$, $c_2 > 0$
   or as $\tau \to  - \infty$   for $U_1 > 0$ and $c_2 < 0$.

    {\bf Milne-type asymptotics.} Now we put $d_1 = 1$. We get
    \beq{5.4}
    \alpha_{+} = (1, 0), \qquad
    \alpha_{-}  = \frac{1}{1 + d_2} (1 - d_2, 2).
    \eeq

    For  $M_1 = \R$, $g^{(1)} = - w dy^1 \otimes dy^1$,
    $ - \infty < y^i < + \infty$, we get a Milne-type (flat)
    asymptotic:

    i) as  $\tau \to + 0$   for $U_1 > 0$ and $c^2 <
    0$;

    ii) as  $\tau \to + \infty$   for $U_1 < 0$ and $c^2
       >  0$.

    Both cases correspond to $u \to + \infty$.

    For  $M_1 = S^1$, $g^{(1)} =  w dy^1 \otimes dy^1$,
    $0 < y^i < + 2 \pi$, we may get either non-singular
    (static) solution in the case i) ($\tau = \rho$) or
    asymptotically flat (static) solution in the case ii).

   \section{Conclusions and discussions}

   In this paper we have considered the exact cosmological type solution with 1-component
   anisotropic fluid. This solution is defined on the
   product manifold (\ref{1.2}) containing  $n$ Ricci-flat factor spaces
   $M_1, ..., M_n$.

   We have singled out a special solution   governed  by the $cosh$ moduli
   function and shown that this solution has Kasner-like asymptotics in the
   limits $u \to \pm \infty$, where $u$ is the harmonic  variable,
   or, equivalently, in the limits $\tau \to  + 0$ and $\tau \to  + \infty$,
   where $\tau$ is the synchronous  type variable.

   We have found a relation between two sets of Kasner parameters
   $\alpha_{\infty}$ and  $\alpha_{0}$. The
   relation between them  $\alpha_{\infty} = S(\alpha_{ 0})$
   is coinciding with the ``scattering law'' formula
   obtained for the $S$-brane solution from   \cite{IM-sl-09}
   when scalar fields are absent and the fluid $U$-vector
   is equal to the brane one.

    The function $S$
   (defined on the set of Kasner vectors obeying $U(\alpha) > 0$) is
   bijective. The inverse function $S^{-1}$
   (defined on the set of Kasner vectors obeying  $U(\alpha) < 0$)
   is given by the same formula as the function  $S$. The
   function $S$ depends upon the co-vector  $U = (U_i)$. It is invariant
   upon the replacement: $U \mapsto \lambda U$, where $\lambda >
   0$ (see \cite{IM-bil-rev09}).
   The transformation $U \mapsto - U$ implies the replacement $S \mapsto
   S^{-1}$.

   We have also analyzed the geometric meaning of the scattering
  law formula in terms of transformation of the Kasner sphere $S^{n-2}$,
  $n \geq 2$. For   $(U,U^{\Lambda}) \neq 0$ (or, equivalently,
  when $\sum_{i = 1}^n U_i \neq 0$)
  we get just a modified inversion with respect to a point $v$
  located outside the Kasner sphere $S^{n-2}$, while for
  $(U,U^{\Lambda}) =0$ (or, equivalently, when $\sum_{i = 1}^n U_i \neq 0$)
  we are led to a reflection with respect to  a hyperplane which
  contains a center of the Kasner sphere.

   The scattering law formula may be applied for the solutions
   with Kasner-like asymptotical behaviours (written
   in a slightly different form)
   \bear{6.a}
   g_{as} = - d\tau \otimes d\tau + \sum_{i=1}^{n} A_i^2
                           (\tau_0 - \tau)^{2 \alpha^i} g^{(i)},
  \ear
  where either $\tau \to \tau_0 - 0$, or  $\tau \to  - \infty$.
  In this case the metric (\ref{6.a}) may describe
  an asymptotical accelerated expansion of flat 3-dimensional factor space
  $M_1$ if $d_1 = 3$,
  $g^{(1)} = dy^1 \otimes dy^1 + dy^2 \otimes dy^2 + dy^3 \otimes dy^3$
  and  $\alpha^1 < 0$.

  Another  application of the scattering law formula  appears when $d_1 =1$
  and one of the asymptotical Kasner set of parameters
  in (\ref{1.m})   is of Milne type:  $\alpha = (1,0,\dots,0)$,
  e.g. when static non-singular solutions ($w = +1$, $M_1 = S^1$)
  or cosmological   solutions ($w = - 1$, $M_1 = \R$) with a horizon
  (for $\tau \to +0$) are considered.
  (Compare with  flux-brane and $S$-brane solutions
   \cite{I-flux,GIM-07}).
  These topics (mentioned above) may be a subject of separate
  publications.


\renewcommand{\theequation}{\Alph{subsection}.\arabic{equation}}
\renewcommand{\thesection}{}
\renewcommand{\thesubsection}{\Alph{subsection}}
\setcounter{section}{0}

\section{Appendix}

\subsection{Solution for Liouville  system}

  Let
 \beq{A.1}
   L=\frac12<\dot x,\dot x> - A \exp[ 2<b,x> ]
 \eeq

be a Lagrangian, defined on $V \times V$, where $V = \R^n$, $A
 \neq 0$,  and  $<\cdot,\cdot>$ is non-degenerate real-valued
quadratic form on  $V$. (Here $\dot x = \frac{dx}{dt}$ etc.)

Let $<b,b>  \neq 0$. Then, the Euler-Lagrange equations for the
Lagrangian (\ref{A.1})
  \beq{A.4}
   \ddot x + 2 A b \exp[ 2<b,x> ] = 0
 \eeq

have the following solution \cite{GIM-95}

 \beq{A.5}
  x(t)= -  \frac{b}{<b,b>} \ln
 |{f}(t-t_{0})| + t \alpha + \beta ,
 \eeq

  where $\alpha,\beta \in V$,
  \beq{A.6}
  <\alpha,b> = <\beta,b> = 0 ,
  \eeq
  and

  \beq{A.7}
 \begin{array}{rlll}
  f(\tau)=
  & R \sinh(\dys\sqrt{\mst C}\tau),& C >0,&
     <b,b> A <0 , \\
 & R \sin(\dys\sqrt{\mst |C|}\tau),&
    C<0,&  <b,b> A <0 , \\
 & R \cosh(\dys\sqrt{\mst C}\tau),& C >0,
  &  <b,b> A >0 , \\
  & |2A <b,b>|^{1/2}\tau,& C=0,&  <b,b> A <0,
\end{array}
\eeq
 where $R =|\frac{2 A <b,b>}{C}|^{1/2}$ and $C$, $t_{0}$ are constants.

  The energy
  \beq{A.8}
   E = \frac12<\dot x,\dot x> + A \exp[ 2<b,x> ]
 \eeq
 calculated for the solution (\ref{A.4}) reads
   \beq{A.8e}
    E = \frac{C}{2<b,b>} + \frac{1}{2}  <\alpha,\alpha>.
  \eeq

\subsection{Lagrange representation}

The Einstein equations (\ref{1.1}) imply the conservation law

  \beq{B.0}
    \nabla_{M}T^{M}_{N}=0.
  \eeq

 that due to relations
(\ref{1.3a}) and (\ref{1.5})  may be written in the following form

  \beq{B.7}
   \dot{\hat{\rho}}+\sum_{i=1}^n
   d_i\dot{\beta^i}({\hat{\rho}}+{\hat p}_i)=0.
  \eeq

  Using the equation
 of state (\ref{1.7}) we get

 \beq{B.7a}
  \kappa^2 {\hat{\rho}}=
  -w A e^{2U_i \beta^i - 2\gamma_{0}},
  \eeq

where $\gamma_0(\beta)= \sum_{i=1}^{n} d_{i} \beta^i$, and $A$ is
constant.

The Einstein equations (\ref{1.1})  with the relations (\ref{1.7})
and (\ref{B.7a}) imposed are equivalent to the Lagrange equations
for the Lagrangian (for $w = -1$ see \cite{IM-95})

 \beq{B.L}
    L = \frac {1}{2}e^{-\gamma+ \gamma_0(\beta)}
    G_{ij}\dot{\beta}^{i}\dot{\beta}^{j}
      -e^{\gamma - \gamma_0(\beta)}V,
  \eeq

where

 \beq{B.32}
  V =  A e^{2U_{i} \beta^i},
 \eeq

 is the
 potential and the components of the minisupermetric $G_{ij}$ are
 defined in (\ref{2.2a}).

 For $\gamma=\gamma_0(\beta)$, i.e. when the harmonic time gauge is
 considered, we get the set of Lagrange equations for the
 Lagrangian

 \beq{B.31}
 L=\frac12 G_{ij} \dot \beta^i \dot \beta^j - V,
 \eeq

 with the zero-energy constraint imposed

 \beq{B.33n}
  E = \frac12G_{ij} \dot
                \beta^i \beta^j + V =0.
       \eeq

  \subsection{The solution}

 The exact solutions for the Lagrangian (\ref{B.31}) with the
potential (\ref{B.32}) could be readily obtained using the
relations from Appendices {\bf A} and {\bf B}.

The solutions read:
 \beq{B.34}
 \beta^i(u)= - \frac{U^{i}}{(U,U)}\ln |f(u)| + c^i u + \bar{c}^i,
 \eeq

where  $u_0$ is constant. Function $f(u)$ in (\ref{B.34}) is the
following:

function reads
 \bear{B.4.5}
  f(u)=
  R \sinh(\sqrt{C}(u-u_0)), \;
  C > 0, \; K A <0; \\ \label{B.4.7}
  R \sin(\sqrt{|C|}(u-u_0)), \;
  C<0, \; K A < 0; \\ \label{B.4.8}
  R \cosh(\sqrt{C}(u-u_0)), \;
  C > 0, \; K A > 0; \\ \label{B.4.9}
  |2A K|^{1/2}(u-u_0), \; C=0, \; K A <0,
  \ear
 where $K = (U,U)$, $R = |2A K/ C|^{1/2}$
 and $C$, $u_0$  are constants.

Vectors $c=(c^i)$ and $\bar c=(\bar c^i)$ satisfy the linear
constraint relations (see (\ref{A.6}) in Appendix {\bf A})

  \bear{B.49}
   U(c)= U_i c^i=0,
  \\ \label{B.50}
  U(\bar c)=  U_i \bar c^i=0.
  \ear

The zero-energy constraint reads (see  (\ref{A.8}) in  Appendix
{\bf A})

 \beq{B.53}
 E= \frac{C}{2(U,U)} + \frac12 G_{ij}c^ic^j=0.
 \eeq

 \subsection{Proof of the inequality (\ref{3.4})}

  Let us prove the inequality (\ref{3.4})

    $$|(s,U^{\Lambda})| >  \frac{|(U^{\Lambda},U)|}{(U,U)} > 0,$$

  for a vector $s = (s^A) \in \R^n$ obeying relations
    $(s,U) = 0$,  $(s,s) =  - 1/(U,U)$.
  Here the scalar-product    $(U,U')= G^{ij} U_i U'_j$,
  where   $G^{ij}= \delta^{ij}d_i^{-1} + (2-D)^{-1}$. We also use here
  the following relations  $(U,U) > 0$, $U^{\Lambda} = (d_i)$ and
   $(U^{\Lambda},U^{\Lambda}) < 0$.

   {\bf Proof.} Let us define the vector
    \beq{D.1}
    U_1 = U -  \frac{(U,U^{\Lambda})}{(U^{\Lambda},U^{\Lambda})}
    U^{\Lambda}.
    \eeq
    It is clear that $(U_1,U^{\Lambda})  = 0$ and
    \beq{D.2}
    (U_1,U_1) = (U,U) -
    \frac{(U,U^{\Lambda})^2}{(U^{\Lambda},U^{\Lambda})}> 0.
    \eeq
     since  $(U,U) > 0$ and  $(U^{\Lambda},U^{\Lambda}) < 0$.
     Let us define vectors:
     \bear{D.3}
     s_0 =  \frac{(s,U^{\Lambda})}{(U^{\Lambda},U^{\Lambda})}
     U^{\Lambda},
    \\ \label{D.4}
     s_1 =  \frac{(s,U_1)}{(U_1,U_1)} U_1,
     \\ \label{D.5}
      s = s - s_0 - s_1.
      \ear
     $s_0$, $s_1$ and $s_2$ are mutually orthogonal
    and hence
    \beq{D.6}
       (s,s) = (s_0,s_0) + (s_1,s_1) + (s_2,s_2).
      \eeq

     For the  first two terms in r.h.s. of (\ref{D.6})
     we get
     \bear{D.7}
       (s_0,s_0) =
       \frac{(s,U^{\Lambda})^2}{(U^{\Lambda},U^{\Lambda})},
       \\ \label{D.8}
       (s_1,s_1) = \frac{(s,U_1)^2}{(U_1,U_1)} =
        \frac{(s,U^{\Lambda})^2}{(U^{\Lambda},U^{\Lambda})}
        \frac{(U,U^{\Lambda})^2}{[(U,U)(U^{\Lambda},U^{\Lambda}) -(U,U^{\Lambda})^2]}
      \ear
       that implies
      \beq{D.9}
       (s,s) =
        \frac{(s,U^{\Lambda})^2
        (U,U)}{(U,U)(U^{\Lambda},U^{\Lambda})
        -(U,U^{\Lambda})^2} + (s_2,s_2).
      \eeq
      For the third term in r.h.s. of (\ref{D.6}) the following
      inequality is valid
       \beq{C.10}
       (s_2,s_2) \geq 0,
      \eeq
       Indeed, due to $(s_2,U^{\Lambda})= 0$, or, equivalently,
       $\sum_{i = 1}^n s_2^i d_i = 0$ , we obtain
       \beq{D.11}
        (s_2,s_2)= G_{ij} s_2^i s_2^j =
        \sum_{i = 1}^n (s_2^i)^2 d_i  \geq 0.
       \eeq
       Using this inequality, (\ref{D.9}),
       $(U^{\Lambda},U^{\Lambda}) < 0$ and $(s,s) =  - 1/(U,U)$
       we get
       \beq{D.12}
        (s,U^{\Lambda})^2 = [ \frac{(U,U^{\Lambda})^2}{(U,U)} -
        (U^{\Lambda},U^{\Lambda})] [(U,U)^{-1} + (s_2,s_2)] >
        \frac{(U,U^{\Lambda})^2}{(U,U)^2} > 0,
      \eeq
       that is equivalent to the  inequality (\ref{3.4}). Thus, (\ref{3.4})
       is proved.

 \begin{center}
 {\bf Acknowledgments}
 \end{center}

  This work was supported in part by the Russian Foundation for
 Basic Research grants Nr. 09-02-00677-a.

\end{document}